# Magnetisation dynamics in the normal and condensate phases of $UPd_2Al_3$

# II: Inferences on the nodal gap symmetry

N. Bernhoeft[1], A. Hiess[2], N. Metoki[3], G. H. Lander[3,4] and B. Roessli[5]

[1] Dépt. de Recherche Fond. sur la Matière Condensée, CEA-Grenoble, F-38054 Grenoble, France

[2] Institut Laue Langevin, BP 156X, F-38042 Grenoble, France

[3] Advanced Science Research Centre, Japan Atomic Energy Research Institute, Tokai, Naka, Ibaraki 319–1111, Japan

[4] European Commission, JRC, Institute for Transuranium Elements, D-76125 Karlsruhe, Germany

[5] Laboratory for Neutron Scattering, ETH Zurich and Paul Scherrer Institute, CH-5232 Villigen, Switzerland

## Abstract:

This paper provides an analysis of neutron inelastic scattering experiments on single crystals of $UPd_2Al_3$. The emphasis is on establishing robust, general, inferences on the joint antiferromagnetic-superconducting state which characterises $UPd_2Al_3$ at low temperatures. A distinction is drawn between these conclusions and various theoretical perspectives of a more model sensitive nature that have been raised in the literature.

## 1.	INTRODUCTION

In this paper the focus is on the inferences that may be drawn from inelastic neutron scattering data on the nature of the antiferromagnetic-superconducting state in $UPd_2Al_3$. In particular, the aim is to establish the scope, and limits, on global properties concerning the symmetry and magnitude of the superconducting energy gap, $\Delta(\boldsymbol{k})$, and the quasiparticle pairing potential from observed changes in spectral form on passing below $T_{sc}$. Extensive reference will hence be made to the experimental evidence presented in $I^{(1)}$ where the results, obtained from samples prepared and measured in independent institutes, point to the robust nature of the thermodynamic physical properties of this material. It is this underlying commonality that forms the backbone of the present analysis and gives credence to the conclusions drawn.

Whilst verification through full, quantitative calculations of the neutron scattering cross section is, to our knowledge, not feasible, it is possible to establish those signatures of the superconducting state that follow from general arguments and to differentiate these from conclusions of a more highly model specific nature. Thus for example, detailed, bandstructure and model dependent results will not be given since such approaches have been extensively discussed elsewhere.[2-10]

In particular it will be seen that whilst unique, wave vector and energy dependent, information on $\Delta(\boldsymbol{k})$ is forthcoming, little can be said on the pairing mechanism of the superconducting state. Indeed, in our present state of knowledge, we feel unable to offer any firm conclusions on this point. Nevertheless, symmetry constraints can be placed that serve as a yardstick against which the various propositions may be measured. To commence, a brief review of the underlying assumptions and constraints implicit in any analysis is given.

## 2.	ANALYSIS OF THE RESULTS

### A	Basic considerations

Long wavelength probes are, to a good approximation insensitive to the translation symmetry operations of the lattice and all periodically related repeat units respond in a similar manner. Conventional optical and microwave spectroscopies, together with transport and thermodynamic measurements of the superconducting state fall in this class. Use of oriented, monocrystalline, samples in conjunction with polarisation techniques may yield directional sensitivity in propitious cases; however, inferences on, for example, the energy gap are limited to its magnitude and crystallographic point group symmetry. The unique role of inelastic neutron scattering as a probe of $\Delta(\mathbf{k})$ lies in its simultaneous wave vector and energy selectivity on the atomic and thermal scales respectively. This sensitivity, to the translation operations of the lattice, brings to light a primary role of the space group symmetry of $\Delta(\mathbf{k})$) and, we argue that in favourable cases, permits one to extract previously inaccessible information.

Experimental results, principally in the form of polarisation analysis and study of the effective form factor[19-22] have established the electronic origin of the anomalous scattering observed below $T_{sc}$. The present work considers the inferences that may be drawn from such data involving the interaction of the neutron with both the condensate and the strongly correlated electronic quasiparticles of the superconducting state. To enable progress, an analysis of the spectral form of the generalised magnetic susceptibility is required.

## B    Generalised magnetic susceptibility

In the following we examine a model dynamical susceptibility wherein the generic approach is to dissect the empirically determined $\chi_q(\omega)$ in to two, or more, distinct components. A similar conceptual fragmentation, used for example in analyses of thermodynamic and μSR data, appears, on occasion, to have been attributed to an assignment of a 'dual' character to the *5f* wave function. However, independent of detailed poles or resonances, there is only one magnetisation de-correlation function. In cases of a multiple-peaked structure observed in frequency and/or wave vector as observed in the response around $\mathbf{Q_0}$ (see part I), care must be taken in any decomposition to preserve the characteristic amplitude and phase correlations of the N-body state.

## B-1    Primitive low frequency-high frequency model

A primitive model for such a structured response is to split the dynamical susceptibility into two distinct components which are simply summed as incoherent contributions to generate a total response function:

$$\chi = \chi_1 + \chi_2 \tag{1}.$$

At this level one may discuss independent contributions, attributed respectively to a low, $\chi_1$, and high, $\chi_2$, energy part of $\chi$. An initial analysis of this form in UPd$_2$Al$_3$, which highlighted the key *qualitative* changes observed on passing below $T_{sc}$, was presented by the JAERI group.[11] The strong renormalisation in $\chi_1$ inferred at $T_{sc}$ evidences the *influence* of superconductivity on the magnetic response function. It is, however, not proof that the superconductivity is driven by the magnetic fluctuations.

## B-2    Low frequency-high frequency coupling model

An ensuing level of sophistication is to take a coupling, generally of mean field form, between the low and high frequency fragments of the full 5$f$-neutron scattering *amplitude* with a fully dynamical (space-time retarded) coupling constant, $\overline{\lambda} = \overline{\lambda}'(q,\omega) + i\overline{\lambda}''(q,\omega)$ in an attempt to restore at least some of the principle correlations of the macrostate. However, in practice, this method is normally approximated by replacing $\overline{\lambda}$ by a constant, $\lambda$, which is used in a direct calculation of the magnetic susceptibility, i.e. calculations at the level of probabilities (scattering cross sections). The field has an abundant literature with many, often equivalent, formalisms.[12] Even this minimal consideration may trigger profound modifications of spectral form. Principally, the effects arise on account of a built in positive feedback giving the net response a Stoner like denominator that acts to enhance, preferentially, the low frequency part with a concomitant renormalisation of the effective low energy line width, as can be seen by the following simple argument.

First, make a conceptual fragmentation of the magnetic system into low and high energy units, designated by $M_1$ and $M_2$ respectively, in which all internal interactions have been included. Then, with a mean field coupling, $\lambda$, form $M_1 = \chi_1 \left[ H + \lambda M_2 \right]$ and $M_2 = \chi_2 \left[ H + \lambda M_1 \right]$ giving the total magnetisation as $M = M_1 + M_2$ and susceptibility,

$$\chi = \frac{\chi_1 + \chi_2 + 2\lambda\chi_1\chi_2}{1 - \lambda^2\chi_1\chi_2}, \qquad (2)$$

where $\chi_1$ and $\chi_2$ are the individual susceptibilities, $\lambda$ the mean field coupling and the primitive model of Eq. (1) is the $\lambda \rightarrow 0$ limit. At low frequencies the real parts of $\chi_{1,2}$ tend to a constant whilst the imaginary parts are proportional to the frequency. This, for the dissipative component of the total susceptibility as $\omega \rightarrow 0$, yields a denominator $1 - \lambda^2 \operatorname{Re}[\chi_1]\operatorname{Re}[\chi_2]$. In contrast, at high frequencies the susceptibilities $\chi_{1,2}$ tend to zero and the denominator goes to unity. Thus, an increase in low frequency response, ultimately driving a divergent response and transition of phase, can be incited through an augmented value of either of $\chi_{1,2}$ and/or the coupling constant. In the interest of simplicity it is often argued, as we do below, to keep the coupling constant, local in space-time and temperature independent. Schematics of the response arising from such general coupling models for the normal antiferromagnetic state in UPd$_2$Al$_3$ are given in the left hand panels of **Fig. 1**.

### B-3 The low energy susceptibility, $\chi_1$, in the superconducting state

The fundamental problem facing any interpretation below T$_{sc}$ is how to partition the scattered intensity between the excitations of the normal and condensate components. In the following we examine a general model of the dynamical susceptibility taking account of the phase coherence of the paired state on its symmetries and amplitude to resolve this dilemma. At the same time it is used to extract unique, wave vector and energy dependent, information on the energy gap function.

The spin susceptibility of excited quasiparticles below T$_{sc}$ is modified by the effects of (i) superconducting phase coherence and (ii) the presence of a gap in the excitation spectrum of the condensate and is calculated, in the following approximation for a singlet ground state, as,[13,14] $\chi_1 = \chi_{qp} + \chi_c$ where the quasiparticle fraction is given by,

$$\chi_{qp}(\mathbf{q},\omega) = \sum_{\mathbf{k}} \frac{1}{2}\left[1 + \frac{\xi(\mathbf{k+q})\xi(\mathbf{k}) + \cos[\Phi(\mathbf{q})]|\Delta(\mathbf{k+q})||\Delta(\mathbf{k})|}{E(\mathbf{k+q})E(\mathbf{k})}\right] \frac{f(\mathbf{k+q}) - f(\mathbf{k})}{\omega - [E(\mathbf{k+q}) - E(\mathbf{k})] + i\Gamma} \quad (3a)$$

and the condensate by,

$$\chi_c(\mathbf{q},\omega) = \sum_{\mathbf{k}} \frac{1}{4}\left[1 - \frac{\xi(\mathbf{k+q})\xi(\mathbf{k}) + \cos[\Phi(\mathbf{q})]|\Delta(\mathbf{k+q})||\Delta(\mathbf{k})|}{E(\mathbf{k+q})E(\mathbf{k})}\right] \frac{1 - f(\mathbf{k+q}) - f(\mathbf{k})}{\omega - [E(\mathbf{k+q}) + E(\mathbf{k})] + i\Gamma}$$

$$- \sum_{\mathbf{k}} \frac{1}{4}\left[1 - \frac{\xi(\mathbf{k+q})\xi(\mathbf{k}) + \cos[\Phi(\mathbf{q})]|\Delta(\mathbf{k+q})||\Delta(\mathbf{k})|}{E(\mathbf{k+q})E(\mathbf{k})}\right] \frac{1 - f(\mathbf{k+q}) - f(\mathbf{k})}{\omega + [E(\mathbf{k+q}) + E(\mathbf{k})] + i\Gamma} \quad (3b).$$

Each element is the summation over the Brillouin zone of a product of a superconducting phase coherence factor and a Lindhard style function. The $\chi_{qp}$ fraction arises from scattering between the quasiparticle levels while $\chi_c$, the condensate fraction, corresponds with the creation and condensation of quasiparticle pairs in neutron energy loss, and neutron energy gain, scattering respectively. The notation is standard: $\xi(\mathbf{k}) = \varepsilon(\mathbf{k}) - \varepsilon_F$ is the quasiparticle energy relative to the normal state Fermi energy and $E(\mathbf{k}) = \sqrt{\xi(\mathbf{k})^2 + |\Delta(\mathbf{k})|^2}$ the quasiparticle excitation energy above the superconducting state. The factor $\Phi(\mathbf{q})$ is the phase difference between $\Delta(\mathbf{k})$ and $\Delta(\mathbf{k+q})$. It may be noted that the coherence function in Eqs. (3) acts in an opposite sense on the normal and condensate contributions to the cross section. This excludes the simultaneous enhancement of the quasiparticle-hole contribution and the condensate fraction to the scattering cross section.

### B-4 Interpretation of the low energy spectra on passing below $T_{sc}$

In this section we discuss how the observed changes to the low energy spectra on entry to the superconducting phase have been understood within alternative scenarios. In this respect, the simultaneous wave vector and energy resolution of the neutron inelastic scattering technique are of particular importance since they yield unique information on the symmetry and magnitude of the superconducting energy gap. As will be seen, the minimal modifications to the response function (as given in Eqs. (3)), implicit in the conventional theory of the phase coherent state, are, in themselves, sufficient to understand the observations.

A key to understanding comes from **Fig. 2** and Fig. 2a(I) which illustrate the scattering intensity around $\mathbf{Q_o}$. Fig. 2a(I) shows how the observed intensity decreases in proportion with $k_B T$ for $T_{sc} < T < T_N/2$ indicating the thermal response of a temperature independent intrinsic susceptibility. The abrupt discontinuity in scattering intensity below $T_{sc}$ to a quantum zero point mode where, both the amplitude, which jumps to approximately twice that expected from the normal state response as shown in Fig. 2, and the generic form, i.e. opening of gap in response with a temperature dependence that no longer follows the simple $k_B T$ law, are strong indicators that the superconducting ground state is having a profound influence on the magnetic density autocorrelation function.

Differences in inference on the physical nature and symmetry of the superconducting state then arise from attributing the measured changes in response on passing through $T_{sc}$ either (i) purely to changes in the two, frequency decomposed, normal state components of $\chi$ and possibly the coupling constant, as schematised in the central panels of **Fig. 1** or, (ii), by taking into account of the role of excitations out of the developing superconducting ground state, as given in the right hand panels.

In the first scenario, the sole contribution to the cross section arises from quasiparticle-hole excitations of the normal state, which are, for $T < T_{sc}$, subject to the phase coherence constraints of the superconducting state. To have significant spectral weight at $\mathbf{Q_o}$ on passing below $T_{sc}$ at low energy transfer would demand, as an almost inescapable requirement, the presence of superconducting nodes commensurate with $\mathbf{Q_o}$ on sheets of the Fermi surface exhibiting a significant density of quasiparticle states. Such a situation is generally energetically unfavourable and is also at variance with available tunnelling data[2,15]. The observation of an *enhanced* spectral weight in Fig. 2 also implies an additional concentration of the response in wave vector, or a frequency amplification process, as discussed above. The 'tuning' options available are that either the spin wave (exciton) undergoes substantial changes in its amplitude, pole and/or damping at these very low temperatures, or the low energy component changes its characteristic amplitude and/or decay rate, and that an appropriate mixture, with or without changes in the (complex) coupling constant λ, is found. Finally, a condition for the 'inelastic' nature of the low frequency response must be added with such a scenario implying a new pole to have been generated in the normal state response function.

Following the second option, the response at low frequencies and lowest temperatures is dominated by the condensate, as schematised in Fig. 1 right hand panel, with negligible contribution from the normal state excitation spectrum. Such a picture is supported by the dramatic fall in heat capacity at temperatures well below $T_{sc}$, signalling a loss of normal state quasiparticle excitations to the susceptibility, due to the opening of a gap on the high state density, strongly correlated sheets of the quasiparticle Fermi surface. The condensate now plays a key role *both* in structuring the magnetic response on account of its phase coherent nature, *and* in supplying an alternative channel of excitation.

In the debate between these two fundamental mechanisms we consider only the response of a singlet condensate since this is the symmetry compatible with the existing array of thermodynamic, transport and tunnelling data [Refs. 15-18 and references therein]. Alternatives, based on spin-triplet pairing wave functions, which do not appear to be supported by thermodynamic and transport measurements are not considered in detail here even though they may be capable of explaining some features of the neutron data[5,16].

The myriad of possible analyses reduce then to a choice between the frameworks (1) and (2) and a discussion is given of each type. The models considered here encompass a mean field coupling of two hypothetical components to $\chi$ with a simple, real, feedback parameter. The fundamental differences that remain are then the assumed dominance by excitations of the normal state both above and below $T_{sc}$ [Refs. 5, 16], whilst in [Refs. 3, 19-22], at the lowest temperatures, the major contribution at low energy transfer is from the condensate. The second point of divergence is the assertion that the nodal symmetry of the superconducting energy gap is considered as being determined purely by the local point group symmetry[5,16], or that explicit account is to be taken of the underlying lattice symmetry and Fermi surface topology[3, 7, 19-22]. These primary choices then dictate the relative amplitude and phase symmetry of both the excitations of the normal state quasiparticle-hole pairs and the excitations out of the condensate that form the basis of further analysis.

To commence an analysis with Eqs. (3) we note that, both the absence of strong thermal dilation effects and significant changes in magnetic moment on passing below $T_{sc}$ indicate that the low energy quasiparticle phasing, intrinsic to the condensed state, does not greatly alter either the lattice or magnetic potentials which determine the Fermi surface.[23] Hence the spatial symmetries of excitation matrix elements, as expressed through the Lindhard functions should not change. In particular, the resonant magnetic wave vector is anticipated to remain constant.[24] The important aspect of the cross section, as contained in a conventional expansion of the susceptibility below $T_{sc}$, is the introduction of an energy gap in the denominator of the Lindhard sum for $\chi_c$. The normal state quasiparticle response does not acquire the corresponding gap and is expected to remain quasielastic in form below $T_{sc}$ [25].

## C symmetry of $\Delta(k)$
### C-1 Spatial symmetry of $\Delta(k)$

The partition of the measured response in $(q, \omega)$ space between normal and condensate excitations is most readily made on examination of the dynamical susceptibility for excitations of minimal energy. A semi-quantitative analysis illustrates the point. First we note that at low temperatures the Pauli principle restricts attention to those excitations of the condensate which involve quasiparticle states lying close to the Fermi surface, $\left(1 - f(\mathbf{k} + \mathbf{q}) - f(\mathbf{k})\right) \approx 1$, with the normal fluid quasiparticle contribution to the bare susceptibility becoming progressively weaker on lowering the temperature. Examination of the phase coherence factor in $\chi$ reveals that, for excitations of minimal energy, where the quasiparticle excitation energies $\xi(\mathbf{k}) = \xi(\mathbf{k} + \mathbf{q}) = 0$, the phase coherence bracket reduces to $1 \pm \cos\left[\Phi(\mathbf{q})\right]$ for the normal and condensate fractions with the upper and lower sign respectively. Hence, for a significant condensate response at wave vector $\mathbf{q}$, $\Delta(\mathbf{k+q})$ must be the negative of $\Delta(\mathbf{k})$ at least over a sizable portion of the zone. From the observation of an *inelastic* and *enhanced* scattering in the superconducting phase of $UPd_2Al_3$ around the antiferromagnetic reciprocal lattice vectors (*i.e.* $\mathbf{q} = \mathbf{Q_o}$ in Eqs. (3)) the inferences are:

(i)       that the dominant contribution arises from the condensate which

(ii)     has a gap, $\Delta$ $(\boldsymbol{k})$, displaying sign inversion on translation by $\mathbf{Q_o}$ over a major part
        of the zone. That is, the observed scattering suggests a spatially *anti-symmetric*
        form of $\Delta$ be taken, $\Delta$ $(\mathbf{k})$ = -$\Delta$ $(\mathbf{k+Q_o})$ [3, 7, 19-22].

The sign difference between the coherence factor for excitations from the normal,
$1 + \cos\left[\Phi(\mathbf{q})\right]$, and condensate, $1 - \cos\left[\Phi(\mathbf{q})\right]$, fractions is of further experimental importance.
In antiferro-periodic symmetry, i.e. $\phi(\mathbf{q}) = \pi$ for $\mathbf{q} = \mathbf{Q_{afm}}$, the coherence factor effectively
*eliminates* normal state scattering at $\mathbf{Q_o}$, so no quasielastic response remains enabling a clear
definition of the inelastic nature of the condensate response, see central and right hand panel
of Fig. 1. In model systems having a jellium, $\mathbf{Q_o} = 0$, or lattice periodic translation symmetry
the phasing enhancement from the coherence factor reverses. The susceptibility
amplification, due to superconducting correlations for wave vectors separated by a reciprocal
lattice unit, then is a maximum for the gapless quasiparticle excitations from the normal fluid
and is small for excitations involving the condensate.

Expanding briefly on this point we recall that wave vectors, $\mathbf{Q_{mag}}$, spanning regions of high
density of electronic states at the Fermi surface yield an enhanced susceptibility, and hence
neutron cross section, through the Lindhard function. In order for experiments to benefit
from this in the identification of $\Delta(\boldsymbol{k})$ below $T_{sc}$, the wave vector of maximal phasing of the
condensate fraction must be commensurate with $\mathbf{Q_{mag}}$. Maximising $1 - \cos\left[\Phi(\mathbf{q})\right]$ requires
$\phi(\mathbf{Q_{mag}}) = \pi$. In a jellium approximation, or the presence of ferromagnetic correlations, this
implies $\Delta(\mathbf{k}) = -\Delta(\mathbf{k+G})$, where $\mathbf{G}$ is a vector of the reciprocal lattice, forcing $\Delta(\boldsymbol{k}) = 0$.
Conversely, for a minimal period of $\Delta(\mathbf{k}) = \Delta(\mathbf{k+G})$, in which case the condensate fraction
gives zero response at the maxima of $\chi$, a condensate may coexist with the ferromagnetic
correlations. In this case the aspect of lattice translation invariance becomes trivial with the
nodal symmetry of $\Delta(\boldsymbol{k})$ being determined by the crystallographic point group. As noted, in
such materials the *normal* quasiparticle contribution is enhanced on passing below $T_{sc}$. This
leads to the possible observation of a modification in the quasielastic intensity and lineshape
on the energy scale of $\Delta(\boldsymbol{k})$. On account of the low energies involved, at the experimental
level this may be seen as a weak change in the (ferro)magnetic Bragg peak intensity. Finally,
in materials where the condensate phasing wave vector is incommensurate with that of the
susceptibility, one anticipates only indistinct signs of the transition below $T_{sc}$.

## C-2 Spectral form of $\Delta(\mathbf{k})$

The condensate response at $\mathbf{Q_o}$, under the constraint of a favourable phase coherence symmetry and within the restriction that we consider only minimal excitation states having $\xi(\mathbf{k}) = 0$, is given by the imaginary part of the Pauli restricted summation $\sum_{k}^{\tilde{}} 1/\left(\omega - 2\left|\Delta_k\right| + i\Gamma\right)$. This is a sum of complex Lorentzian amplitudes centred at $2\left|\Delta_k\right|$ and of width $\Gamma$. In the case that $\Delta(\mathbf{k}) = -\Delta(\mathbf{k+Q_o})$ with $\left|\Delta_k\right|$ independent of $\mathbf{k}$, i.e. the square wave representation of the antiferro-periodic nodal gap state (previously referred to as the 'antiferromagnetic-*s*-wave' state[3]) the response simplifies to a single pole centred at $2\Delta$ of damping, $\Gamma$, related to the effective quasiparticle lifetime. The presence of a sizeable gap anisotropy, explicit in some models of the superconducting state[5,16] would lead to interference over a sum of complex amplitudes giving a spectral form spread in energy and of diminished magnitude. The sharp profile of the condensate response in energy transfer observed in UPd$_2$Al$_3$ thus appears to favour the conjecture of an isotropic crystallographic point group symmetry of the energy gap over extended *s* or *d*-wave variants[2-10,16,19-22] although detailed calculations are required in each case.

## C-3 Spin symmetry of $\Delta(\mathbf{k})$

The deduced symmetry of the measured gap function, $\Delta(\mathbf{k}) = -\Delta(\mathbf{k+Q_o})$, implies $\Delta(\mathbf{k}) = \Delta(\mathbf{k+G})$. As a consequence $\Delta(\mathbf{k})$ follows the translation symmetry of the crystallographic Brillouin zone, with the inference that the condensate wave function is not sensitive to the magnetic potential that defines the antiferromagnetic unit cell doubling along the hexagonal axis for the spin polarised quasiparticles. This is, *a posteriori*, consistent with the spin pairing symmetry underlying Eqs. (3), and in agreement with available experimental data and detailed calculations based on the computed Fermi surface, that the gap function is a spin singlet[2,7,17,18].

Nevertheless, theoretical approaches differ on this issue. Identifying three separate mechanisms of condensation based respectively on, phonon exchange,[27] spin fluctuation exchange[2,4,26] or a novel crystalline electric field (CEF) exciton mode, as proposed in Sato *et al.*[16] and Thalmeier[5], the results are that:

i)      a phonon driven condensate will have an associated spin singlet (even parity) state on account of the Pauli principle and spin invariance under phonon exchange unless higher order spin-orbit driven effects are invoked. The resulting energy gap translation symmetry is that of the crystallographic Brillouin zone.

ii)     the spin fluctuation condensate may achieve either spin singlet or spin triplet states on account of rotational invariance with a gap translation symmetry periodic in the crystallographic Brillouin zone.

iii)    the CEF-exciton mode of Sato et al.[16] and Thalmeier[5] is of odd parity (i.e. spin triplet symmetry) on account of coupling the Cooper spin pair with a local magnetic moment. Cast within the framework of the translation symmetry of the antiferromagnetic unit cell, the model is obliged to place the nodal region in the equatorial plane. In order to satisfy the observed antiferromagnetic repeat wave vector $\Delta(\boldsymbol{k})$ then requires the orthogonal phase symmetry, i.e. $\Delta(\boldsymbol{k}) \sim \sin(ck_z)$, to that invoked above.

From (i) and (ii), and as a rather robust and general conclusion highlighted by Oppeneer and Varelogiannis[7] in their detailed calculations based on the computed Fermi surface in UPd$_2$Al$_3$, the observed symmetry of the gap function is *not* universally tied to any particular pairing mechanism. Thus, for example, Oppeneer and Varelogiannis show that either phonon or spin fluctuation pairing are capable of producing both *s*-wave and *d*-wave condensates with the self-consistent computed gap in both cases following the symmetry of the crystallographic Brillouin zone in agreement with the general comments given above.

In a highly original analysis used to interpret a model cross section by Sato et al.[16], and made explicit by Thalmeier[5], the basic ingredient is a fragmentation of the *5f* shell into a local moment and itinerant state, with the local moment dynamics described as an excitation of coupled, ionic, CEF levels. This yields a superconducting state of odd parity (i.e. spin triplet symmetry) on account of coupling the Cooper spin pair with the magnetic moment, a conclusion apparently in contradiction with the inferences of available thermodynamic and transport data. A central role of the exciton mode in the vicinity of $\mathbf{Q_o}$ is ensured by the condition of its accidental degeneracy with the magnitude of the superconducting energy gap at each $T < T_{sc}$. This assumption appears to require both a substantial superconducting energy gap maximum at the $\mathbf{Q_o}$ point in the zone and a change in nature of the normal state response below $T_{sc}$, apparently at variance with, on the one hand, interpretation of tunnelling measurements[2,15] and, on the other, the implications of Eqs. (3). Additionally, since the neutron response arising from the condensate is given by a Lindhard sum over all $|\Delta(\boldsymbol{k})|$, see Eqs. (3) and following discussion (section C2), the implied variance of $|\Delta(\boldsymbol{k})|$ would be expected to result in a broad, weak, spectral form as opposed to the strongly enhanced resonance observed in Fig 2. The softening of the zone centre CEF level (exciton) on entering the superconducting phase required to generate sufficient feedback enhancement in the low frequency mode of ~ 30 % is certainly dramatic evidence of $T_{sc}$ in this scenario given it is taken to represent the stable, (quasi)-localised, $5f$ component of the response. Further, the fundamental assumption of strongly localised $5f$ levels is difficult to reconcile with the lack of observation of CEF levels in the paramagnetic state[28] and the success of *ab-initio* band-structure calculations using the delocalised LSDA approach to reproduce experimental Fermi surface areas, as measured by de Haas-van Alphen effect, which often are taken as an indication that the $5f$ levels are largely delocalised.[29,30]

## D Tunnelling, Fermi surface topology and |Δ (k)|

Interpretation of tunnelling data underlines the importance of the space group symmetry of the energy gap. When the symmetry of the energy gap is tied to the lattice, as is the structure of the Fermi surface, failure to account for the simultaneous constraints may lead to assignment of alternative symmetries to the energy gap to rationalise tunnelling, transport and thermodynamic data. As the neutron scattering response reveals, in $UPd_2Al_3$ the lattice symmetry appears to be determinant, enforcing nodes of the energy gap periodic within the Brillouin zone. The additional constraint of Fermi surface topology, on the optimisation of the condensation free energy below $T_{sc}$, favours an energy gap with nodal structure in the vicinity of low density quasiparticle states. A conclusion corroborated in $UPd_2Al_3$ by the rapid decrease of heat capacity below $T_{sc}$ [17, 18]. Examination of the band structure in $UPd_2Al_3$ shows the antiferro-periodic nodal gap state to obey this criterion with the high state density 'egg' Fermi surface sheet,[29] as identified in both inelastic neutron scattering[19-22] and tunnelling experiments,[2,15] lying close to, but inside the Brillouin zone boundary.[3]

The available measures of |Δ|, from both tunnelling and inelastic neutron scattering at the comparable point in the zone, thus support a lattice symmetry determined gap function with nodes along *c\**. In other words, with minimal assumption it appears unnecessary to invoke a *d*-wave point group character of the energy gap. Indeed, the totally symmetric, $A_{1g}$, *s*-wave point group, in conjunction with the Fermi surface as calculated, and substantiated through independent dHvA measurements,[30] is able to account not only for the very good correlation in magnitude of gap along the *c\** axis as determined by both tunnelling and

neutron spectra, but also for the observation of many apparently anomalous thermodynamic and transport properties in this material.[3] As a corollary to these observations, the antiferro-periodic nodal gap scenario suggests a discontinuous sensitivity of the superconducting state to applied pressure. On expansion of the Fermi surface a catastrophic collapse of the condensate energy as the nodal planes at the antiferromagnetic Brillouin zone boundary are approached will lead to a rapid destruction of $T_{sc}$ at a critical pressure. Concomitant with this eventuality, the fusion of the approaching Fermi surface 'eggs' on either side of the antiferromagnetic Brillouin zone boundary signals the collapse of the antiferromagnetic potential and hence supports published data that suggests the simultaneous collapse of both the magnetic and superconducting order parameters[31]. A direct test by dHvA measurements may be technically feasible[32].

**E Coupling schemes for superconductivity**

Whilst valuable new information on the energy gap magnitude and symmetry is available, in the absence of quantitative calculations for the cross section the use of inelastic neutron scattering to determine the coupling mechanism of the condensed state is speculative. As a basis for discussion, one may ask such exercises to (i) yield a parity of the proposed superconducting paired state consonant with that inferred from thermodynamic and transport data, and (ii) offer a rationale for the appearance of an energy gap with (iii) reinforced quantum excitation below $T_{sc}$ in the inelastic neutron scattering spectra at (iv) *one* select point in the Brillouin zone, i.e. of defined translation symmetry, as demonstrated by the extensive mappings of $\chi(q,\omega)$ presented in I. These four basic measures may then serve as a yardstick against which to compare the strikingly different model coupling mechanisms presented[2,4-10].

**3.      CONCLUSION**

The purpose of this paper has been to present the minimal requirements of any analysis of the low temperature magnetic inelastic response in $UPd_2Al_3$. Inferences on the renormalisation of electronic correlations in the superconducting state are made on the basis of differences in neutron scattering spectral weight on passing below $T_{sc}$. Crucially, in favourable materials such as $UPd_2Al_3$, the changes are of sufficiently distinct character as to allow partition of the observed intensity into its condensate and normal fluid fractions. In any approach, the critical influence of phase correlations which characterise the superconducting ground state on excitations of *both* the normal and condensate fractions is noted.

In $UPd_2Al_3$ the observed differences permit robust inferences of the global gap symmetry which embody the lattice translation invariance in addition to estimates of its magnitude at select points in the Brillouin zone. In this context it is of interest to note that the increasing range of antiferromagnetic quasielastic correlations to ~ 100 Å on cooling to $T_{sc}$, Figs. 3(I), 4(I), which approaches the estimated coherence length of the condensate[17,18] suggests the passage to the superconducting critical point may be foreshadowed in the response of the normal phase. When below $T_{sc}$, the quasielastic scattering vanishes to be replaced by the gapped condensate response, Fig. 2. It is this enhanced, resonant, spectral feature which yields microscopic information on the energy and, unique to neutron scattering, the wave vector dependence of the symmetry of the superconducting energy gap. These observations suggest the full group symmetry of the energy gap to be basic to an understanding of the thermodynamic and transport properties observed in this material[3].

On the other hand, the neutron data have little information content on the detailed pairing mechanism itself[7] such inferences rapidly becoming highly model sensitive both on account of theoretical approximation schemes adopted and numerical complexity in calculation. The situation is further aggravated in cases where a multi-modal nature of $\chi(\omega)$, with the inherent need for some decoupling approximation,[3,16,19-22] gives parametric uncertainty in pole positions and widths, which adds to the intrinsic uncertainty of counting statistic limited spectra implicit in the use of a weakly interacting probe.

A further experimental aspect is that the typical superconducting coherence volume needs to be smaller than that of the neutron probe to see the coherence effects implicit in Eqs. (3). When this condition is not fulfilled, the coherent scattering amplitude senses only single quasiparticle like properties. Conversely, to observe a resonant spectral form in energy transfer, the time coherence of the phased, momentum coupled, electronic states should be greater than the probe's temporal decoherence interval. If this is not the case, the inherent width and weak amplitude of spectral response may conspire to make identification uncertain.

Additional information of importance on the nature of the energy gap and pairing mechanism may, however, be forthcoming from experiments below $T_{sc}$ as a function of applied field. Initial investigations[11], at 0.4 K and $\mathbf{Q_o}$ for fields up to 4 T, have revealed that the inelastic pole associated with the superconducting condensate is progressively quenched in amplitude at fixed frequency even for modest fields (B $\leq$ B$_{c2}$/2). This appears at variance with its behaviour on heating in zero field, where up to T$_{sc}$/2 it appears stable in amplitude and form[3].

Whilst considerable progress has been made, developing insights on the nature of the coherent antiferromagnetic-superconducting state remains a major challenge in condensed matter physics. The unique combination of energy selectivity and microscopic spatial nature of inelastic neutron scattering has brought to light the underlying space group symmetry of the energy gap parameter, a feature lost in the spatial averaging of probes such as optical and tunnelling spectroscopies. Further progress will rely on progress in experimental technique, discoveries of other model systems and a deeper understanding of the generic problem of coupled order parameters. In the interim, we hope this basic analysis of the neutron inelastic scattering experimental data on UPd$_2$Al$_3$ will stimulate further investigations. In particular, detailed calculations of the response function both in the normal and antiferro-periodic nodal gap states of this fascinating material are called for.

## 4.      ACKNOWLEDGEMENTS

We thank all colleagues who have helped this work, in particular the critique of P. Oppeneer and advice on Ref. 16 by N. K. Sato and P. Thalmeier is appreciated. Both GHL and NB thank the Director and staff of the Advanced Science Research Centre, JAERI, for warm hospitality during visits that have advanced this collaboration and AH thanks colleagues at IFP, Technische Universität, Dresden, for hospitality during his visit.

**Figure Captions**

**Fig. 1 (colour online)**: *Left hand panel:* schematic of response for $T > T_{sc}$ in the normal antiferromagnetic phase. Top frame illustrates the characteristic bi-modal scattering of $UPd_2Al_3$ which is shown in the middle two frames as a conceptual fragmentation into a coupled low and high energy response. The bottom frame is a cartoon representation of the response (black arrow) of the normal quasiparticle states (red hoops) to the neutron probe (green arrows). The cross section is given in each case by the summation of such events over the space-time scattering volume.

*Central panel:* top frame is a schematic of response for $T < T_{sc}$ in the superconducting antiferromagnetic phase, shown in the middle frames as a conceptual fragmentation into a coupled normal quasiparticle low and high energy response. The vanishing magnitude of the quasielastic response (zero amplitude in a state of antiferro-periodic nodal gap symmetry at low frequency) occurs on account of both the gapping of the Fermi surface, with concomitant loss of low energy excitations as evidenced by the falling heat capacity, and the phase cancelling role of the nodal symmetry of the condensate on such excitations in the magnetic susceptibility, Eqs. (3). Bottom frame is a schematic of the dynamic equilibrium between the paired and Fermion states. Normal quasiparticles, above the energy gap, marked as red hoops and paired state, below energy gap, designated as bound, overlapping hoops. The phase coherent condensate influences the possible excitation processes of the normal state quasiparticles and such interference effects must be taken into account below $T_{sc}$ in analysis of the quasiparticle spectra (designated by yellow mesh).

*Right hand panel:* schematic of response for $T < T_{sc}$ in the antiferromagnetic superconducting phase, shown in lower frames as conceptual fragmentation into a coupled condensate low, and normal quasiparticle high, energy response. The enhanced magnitude of the condensate response occurs on account of phasing role of the antiferro-periodic nodal gap symmetry on such excitations in the magnetic susceptibility. Bottom frame is a schematic of the dynamic equilibrium between the paired and Fermion states as in central panel. As illustrated, direct excitation out of the condensate may occur. Again, the superconducting energy gap has a profound influence on the intensity of scattering at a given momentum and energy transfer, reflecting the symmetry and coherence of the wave function.

**Fig 2:** The scattering observed at $\mathbf{Q_o}$ in the superconducting state (T = 150 mK) as experimental points. Above $T_{sc}$, the low energy spectrum appears as a quasielastic response together with a higher energy, spin wave like, feature. The solid line is smooth fit to 2.5 K data (T > $T_{sc}$) scaled by the Bose factor (with a constant background subtracted). The horizontal bar indicates the instrumental resolution. Data taken on IN14 with $k_f$ = 1.15 Å$^{-1}$.

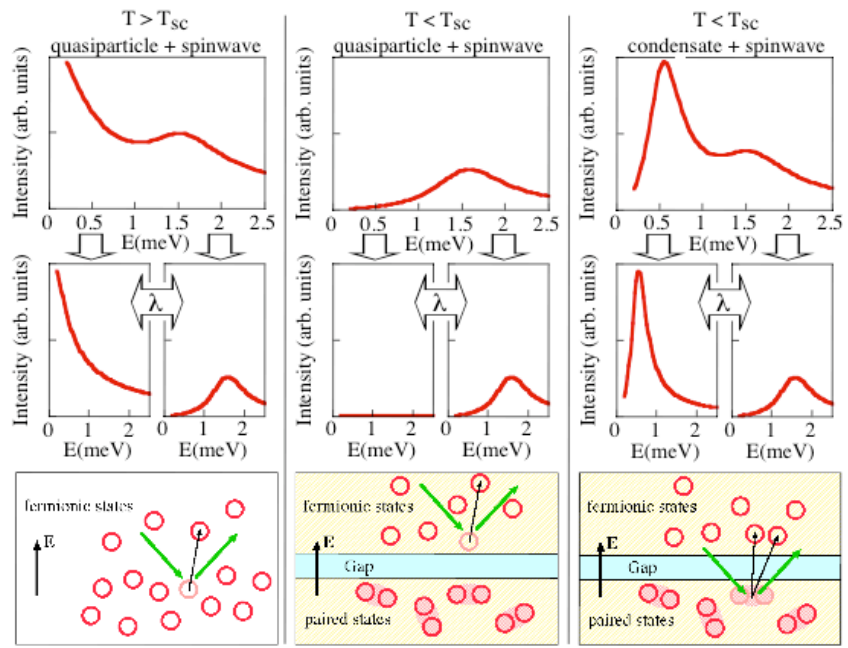

Fig. 1

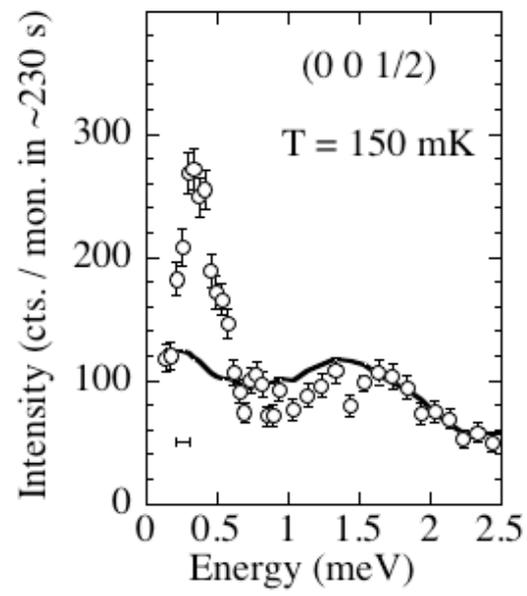

Fig. 2